\documentclass[10pt,twocolumn,twoside] {IEEEtran}
\usepackage{amsmath,epsfig}

\begin{document}

\title{Instantaneous frequency estimation using the discrete linear chirp transform and the Wigner distribution}

\author{Osama A. Alkishriwo~
         and~Luis F. Chaparro

\thanks{Manuscript received ....}}

\markboth{IEEE SIGNAL PROCESSING LETTERS}
{ \MakeLowercase{\textit{ALKISHRIWO et al.}}: Instantaneous frequency estimation using the discrete linear chirp transform and the Wigner distribution}

\maketitle

\begin{abstract}
In this paper, we propose a new method to estimate  instantaneous frequency using a combined approach based on the discrete linear chirp transform (DLCT) and the Wigner distribution (WD). The DLCT locally represents a signal as a superposition of linear chirps while the WD provides  maximum energy concentration along  the instantaneous frequency in the time-frequency domain for each of the chirps. The developed approach takes advantage of the separation of the linear chirps given by the DLCT, and that for each of them, the WD  provides an ideal representation.  Combining the WD of the linear chirp components, we obtain a time-frequency representation free of cross-terms that clearly displays the instantaneous frequency. Applying this procedure locally, we obtain an instantaneous frequency estimate of a non-stationary multicomponent signal. The proposed method is illustrated by simulation. The results indicate the method is efficient for the instantaneous frequency estimation of multicomponent signals embedded in noise, even in cases of low signal to noise ratio.
\end{abstract}

\begin{IEEEkeywords}
Instantaneous frequency, discrete linear chirp transform, time-frequency analysis, Wigner distribution, estimation
\end{IEEEkeywords}

%

\section{Introduction}

\IEEEPARstart{I}{n} many applications in biomedicine, speech processing, communications, radar, underwater acoustics, where non-stationary signals are present, it is typically necessary to estimate the instantaneous frequency of the signals  \cite{Boashash92}.
Time-frequency distributions (TFDs) are widely used for IF estimation based on peak detection techniques \cite{Sejdic08}, \cite{Stankovic03}, \cite{Djurovic11}. The most frequently TFD used for linear chirps is the Wigner distribution (WD) due to its ideal representation for such signals. However, in the case of multicomponent signals, Wigner distribution does not perform well because of the presence of extraneous cross-terms.

Recently, the discrete linear chirp transform (DLCT) \cite{Alkishriwo12} was introduced as
an instantaneous--frequency frequency transformation, capable of locally representing signals in terms of linear chirps. It generalizes the discrete Fourier transform and has an instantaneous--frequency time dual transform, and very importantly it can be efficiently implemented using  the fast Fourier transform (FFT).

The work of \cite{Hussain02} in multicomponent signal IF estimation requires to have a TFD that has high resolution and is free of cross-terms. In \cite{Akan01} an iterative method is proposed for IF estimation using the evolutionary spectrum. In general, the instantaneous frequency estimation requires  signal separation, for multicomponent signals, and high resolution time-frequency distributions.  In this paper, we propose a new method that takes advantage of the DLCT for signal separation, and of the WD for high resolution in the time frequency space.

\section{The Discrete Linear Chirp transform (DLCT)}
 Given a discrete-time signal $x(n)$, with finite support $0\le n \le N-1$, its discrete linear chirp transform (DLCT) and its inverse are \cite{Alkishriwo12}
\begin{eqnarray}
&& X(k,\beta)=\sum_{n=0}^{N-1} x(n) \exp\left( -j\frac{2\pi}{N} (\beta n^2 +kn) \right)
\nonumber \\
&& x(n) = \sum_{\ell=-L/2}^{L/2-1} \sum_{k=0}^{N-1}\frac{X(k,\beta)}{LN} \exp\left( j\frac{2\pi}{N}(\beta n^2+kn)\right)\nonumber\\
&&~~~~~ \qquad 0\le n, k\le N-1
\label{eq3}
\end{eqnarray}
The DLCT decomposes a signal using linear chirps
\begin{eqnarray*}
\phi_{\beta,k} (n)=\exp\left(j\frac{2\pi}{N} (\beta n^2 +kn)\right)
\end{eqnarray*}
characterized by the discrete frequency $2\pi k/N$, and a chirp rate $\beta$,  a continuous variable connected with  the instantaneous frequency of the chirp:
\begin{eqnarray*}
IF(n,k) =\frac{2\pi }{N } (2\beta n+ k).
\end{eqnarray*}
Assuming a finite support for $\beta$, i.e., $-\Lambda\le \beta < \Lambda$, it is possible to construct an orthonormal basis $\{ \phi_{\beta,k}(n)\}$ with respect to $k$ in the supports of $\beta$ and $n$. To obtain a discrete transformation, we approximate the chirp rate as
\begin{eqnarray*}
 && \beta \approx  \ell C, ~~\mbox{where} ~~C=\frac{2\Lambda}{L} ~~\mbox{so that }\\
 && -\frac{L}{2} \le \ell \le  \frac{L}{2}-1~~\mbox{ is integer}.
\end{eqnarray*}
The DLCT is a joint instantaneous-frequency frequency transform that generalizes the discrete Fourier transform (DFT); indeed $X(k,0)$ is the DFT of $x(n)$. Thus, the DLCT can be used to represent signals that locally are combinations of sinusoids, chirps, or both.

It is important to remark that in a discrete chirp, obtained by sampling a continuous chirp satisfying the Nyquist criteria, the chirp rate $\beta$ cannot be an integer.  Indeed, if a finite support continuous chirp
 $$x(t)= e^{j (\alpha t^2 +\Omega_0 t)}\qquad \alpha=\frac{\Delta \Omega}{2\Delta t}, ~~0\le t\le T$$
 is sampled using a sample frequency
 $$\Omega_s=\frac{2\pi}{T_s}= M\Omega_{max}, ~~~M\ge 2,$$
 as determined by the Nyquist criteria, the obtained discrete signal is
 \begin{eqnarray*}
 x(n)&=&  e^{j([\alpha T_s^2] n^2+ [\Omega_0 T_s] n)}\\
 &=& e^{ j(\hat{\beta} n^2+\omega_o n)}\qquad 0\le n\le \frac{T}{T_s}=N-1
 \end{eqnarray*}
where we let $\hat{\beta}=\alpha T_s^2$ be the chirp rate and $\omega_0=\Omega_0 T_s$ be the discrete frequency.
Then the modulated chirp
\begin{eqnarray*}
&&x(n)e^{-j\omega_0 n}=e^{ j\hat{\beta} n^2}~~\mbox{where} \\
&& \hat{\beta}=\alpha T_s^2= \left[ \frac{\Delta \Omega}{2\Delta t}\right]T_s^2 =\frac{\Omega_s T_s/M}{2N}=\frac{\pi/M}{N}
\end{eqnarray*}
therefore,
$$\beta=\frac{N \hat{\beta}}{2\pi}=\frac{1}{2M}$$
is not an integer for $M\ge 2$. Therefore, for not aliased chirps, we need $|\beta|\le 0.25$.

For each value of $\beta$  it can be shown that
$$x_{\beta}(n)=\sum_{k=0}^{N-1}\frac{X(k,\beta)}{N} \exp\left( j\frac{2\pi}{N}(\beta n^2+kn)\right)$$
equals $x(n)$ so that the inverse DLCT is the average over all values of $\beta$.

\section{Instantaneous frequency estimation}
In this section, we introduce a procedure that combines the DLCT and the WD to estimate the IF.  Locally, the DLCT approximates the signal as a sum of linear chirps, for each
of which the WD provides the best representation.  Superposing these WDs we obtain an estimate of the overall instantaneous frequency of the signal.

The Wigner distribution of a signal $x(t)$ is given by \cite{Cohen95}
\begin{eqnarray*}
W(t,\Omega)=\frac{1}{2\pi}\int x^\ast\left(t-\frac{\tau}{2}\right)x\left(t-\frac{\tau}{2}\right)e^{-j\tau\Omega}\mathrm{d}\tau
\label{eq1}
\end{eqnarray*}
And for a linear chirp $x(t)=\exp(j(\alpha t^2/2+\Omega_0 t))$ with instantaneous frequency
\begin{eqnarray*}
IF(t)=\alpha t+\Omega_0
\label{eq2}
\end{eqnarray*}
 its Wigner distribution is
\begin{eqnarray}
W(t,\Omega)=\delta(\Omega-[\alpha t+\Omega_0])=\delta(\Omega -IF(t))
\label{eq1}
\end{eqnarray}
Thus the Wigner distribution of a linear chirp concentrates the energy exactly along the instantaneous frequency in an optimal way. However,  the IF is only clearly seen when the signal is a single chirp, additional terms --- cross-terms --- appear when the signal is composed of more than one chirp.

\begin{figure}[h]
\begin{minipage}[b]{1\linewidth}
\vspace{-1.5cm}
\begin{center}
\includegraphics[trim= 3cm 4cm 2cm 7cm, clip, width=9.5cm]{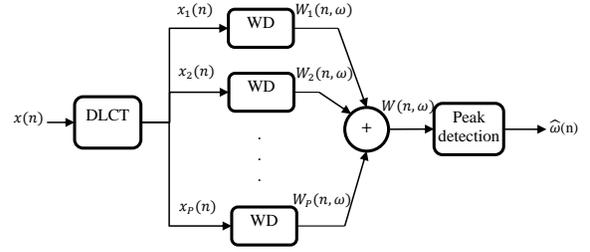}\\
\vspace{-4.0cm}
\end{center}
\end{minipage}
\caption{ Instantaneous frequency estimator.}
\label{fig1}
\end{figure}

If the signal $x(n)$ is the input to the system shown in Fig. \ref{fig1}, the output of the DLCT will be approximated by a sum of linear chirps. Therefore, we can find the WD of each of these linear chirps and synthesize them to obtain a WD free of cross-terms.  Assuming that $x(n)$ is approximated using the DLCT as
\begin{eqnarray}
&& x(n)\approx \sum_{i=1}^{P} x_i(n) ~~~\mbox{where}~~\nonumber\\
&&  x_i(n)= a_i~ \exp\left(j\frac{2\pi}{N}(\beta_i n^2+k_i n)\right)
\end{eqnarray}
and $P$ is the number of chirp components.  The WD of each chirp is given by
\begin{eqnarray*}
 \mbox{W}_i(n,\omega)=\alpha_i\:\delta(\omega-\beta n-\omega_i)
\end{eqnarray*}
where
\begin{eqnarray*}
 \omega_i=\frac{2\pi k_i}{N},\:\alpha_i=a_i^2,~~ \mbox{and}~~ i=1,2,\cdots, P
\end{eqnarray*}
Adding the W$_i(n,\omega)$  we obtain an approximation of the Wigner distribution
 W$(n,\omega)$ corresponding to $x(n)$, but free of cross-components.
Since the Wigner distribution concentrates the energy along the instantaneous frequency, the IF is estimated by
\begin{eqnarray}
\hat{w}(n)= \arg \left [\max\:\: \sum_{i=1}^P \mbox{W}_i(n,\omega) \right ]
\end{eqnarray}

As indicated above, the instantaneous frequency is approximated locally by linear chirps. Thus the signal in general is windowed before applying the above procedure locally. The estimated IF  $\hat{\omega}(n)$ is obtained from the peak detection approach for the high resolution time-frequency distribution which is a result of combining the DLCT with the DW. The accuracy of the estimation is measured by the mean square error
\begin{eqnarray}
 M\!S\!E=\langle \left\{\omega(n)-\hat{\omega}(n)\right\}^2\rangle
\end{eqnarray}
where $\langle . \rangle$ is the average.

\section{Simulations}
To evaluate the performance of the proposed instantaneous frequency estimation method, we consider multicomponent signals with linear, quadratic, and sinusoidal instantaneous frequencies. Also, we add noise to the signals and test our procedure for several signal to noise ratios (SNRs) values.

\begin{figure}[h]
\begin{minipage}[b]{0.5\linewidth}
\vspace{-0.5cm}
  \centering
 \includegraphics[trim= 3cm 7cm 3cm 6.5cm, clip, width=5cm]{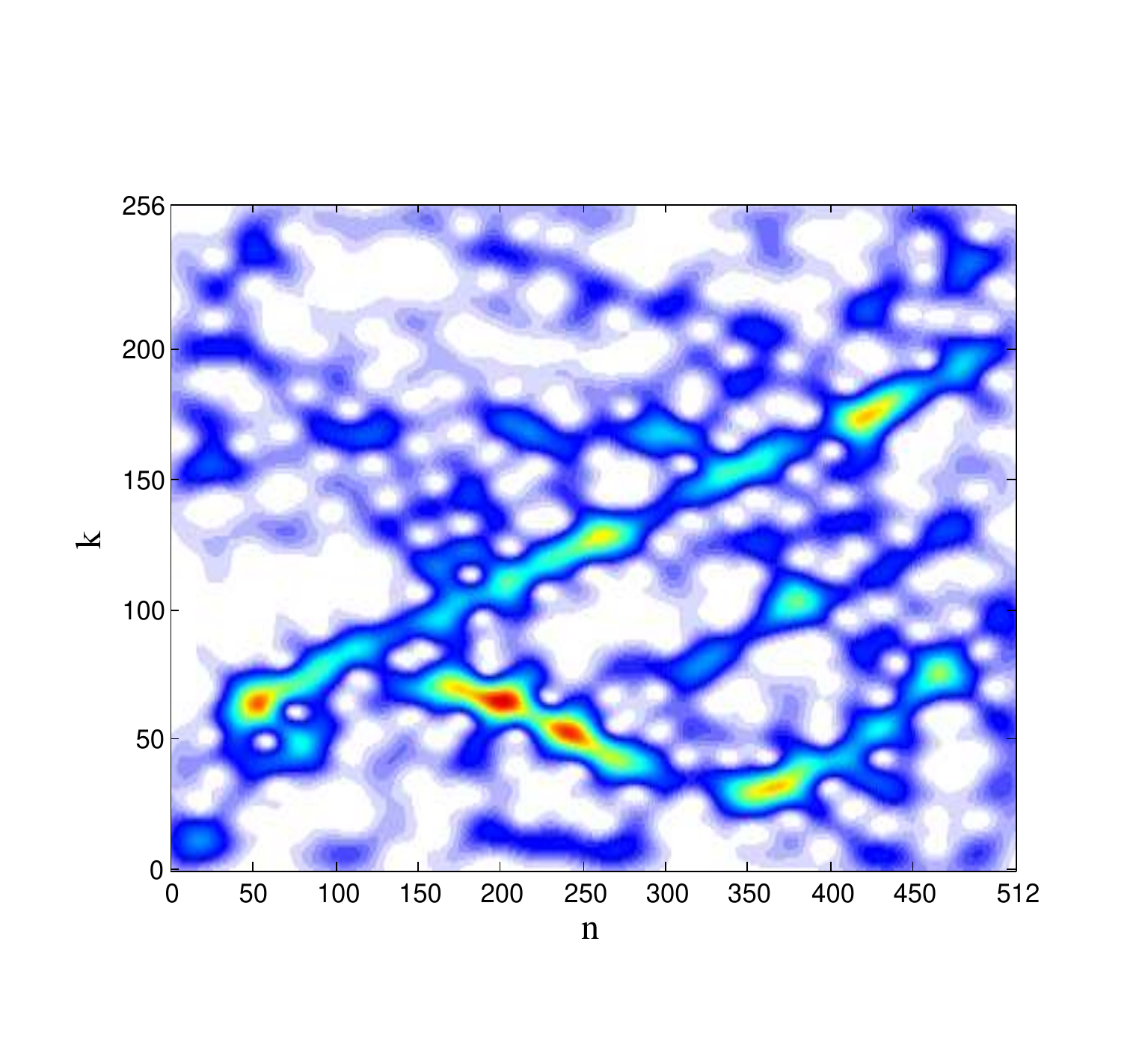}\\
 \vspace{-0.5cm}
  \centerline{(a)}
  \end{minipage}%
  \begin{minipage}[b]{0.5\linewidth}
  \vspace{-0.5cm}
  \centering
  \includegraphics[trim= 3cm 7cm 3cm 6.5cm, clip, width=5cm]{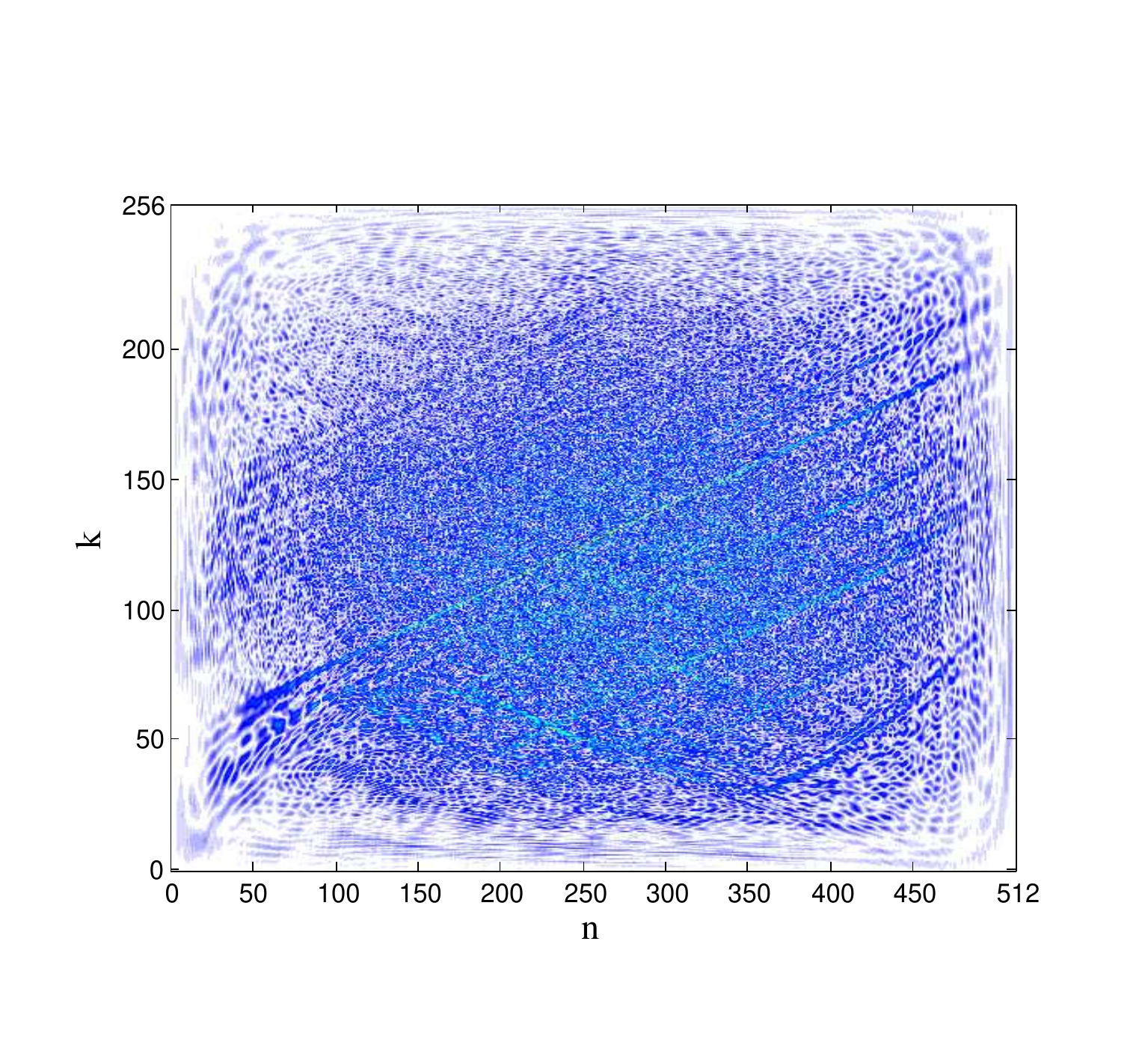}\\
  \vspace{-0.5cm}
  \centerline{(b)}
 \end{minipage}
 \begin{minipage}{0.5\linewidth}
 \vspace{-0.5cm}
\centering
\includegraphics[trim= 3cm 7cm 3cm 6.5cm, clip, width=5cm]{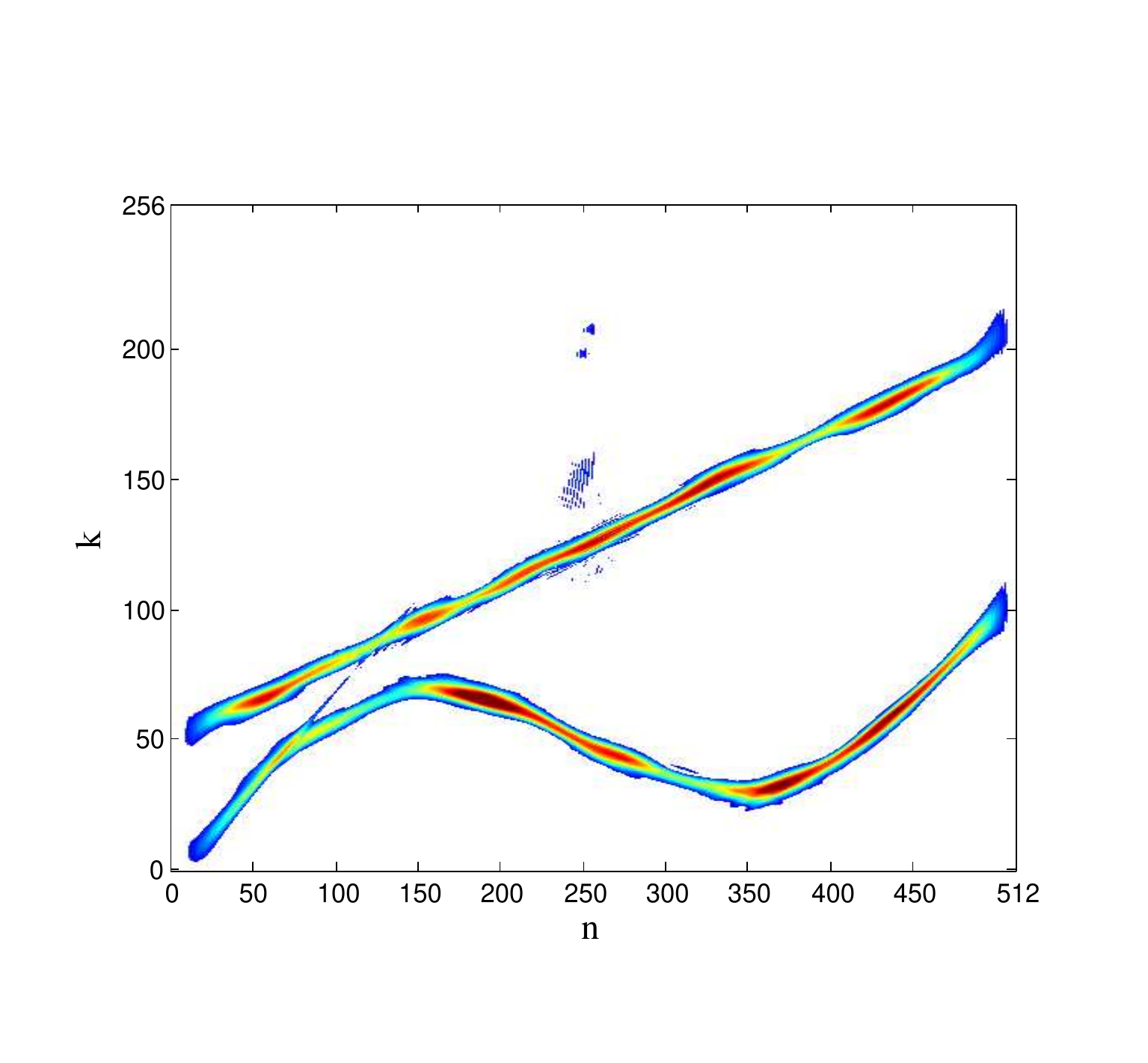}\\
\vspace{-0.5cm}
\centerline{(c)}
\end{minipage}%
\begin{minipage}{0.5\linewidth}
\vspace{-0.5cm}
\centering
\includegraphics[trim= 3cm 7cm 3cm 6.5cm, clip, width=5cm]{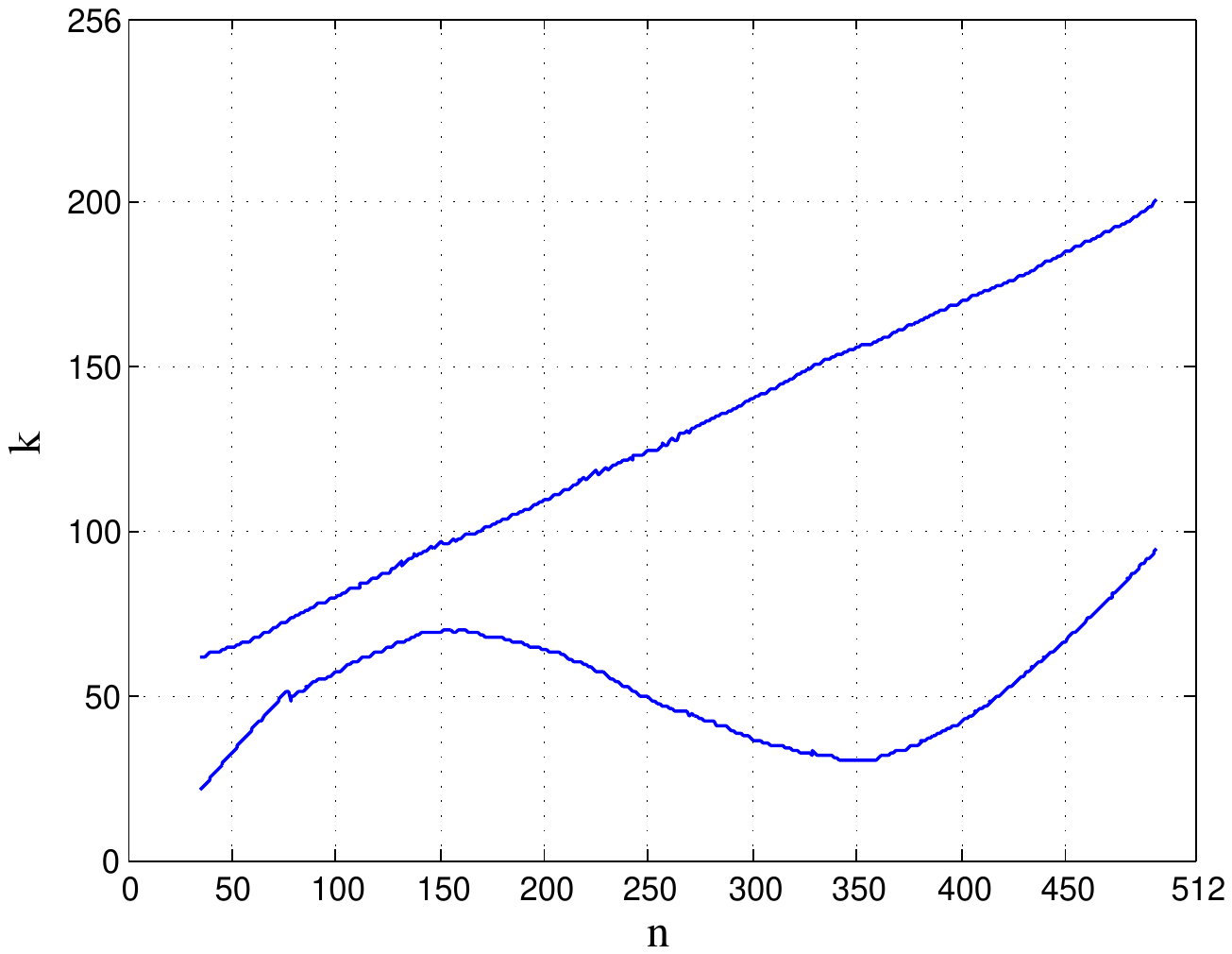}\\
\vspace{-0.5cm}
\centerline{(d)}
\end{minipage}
 \begin{minipage}[b]{0.5\linewidth}
  \vspace{-0.5cm}
  \centering
  \includegraphics[trim= 3cm 7cm 3cm 6.5cm, clip, width=5cm]{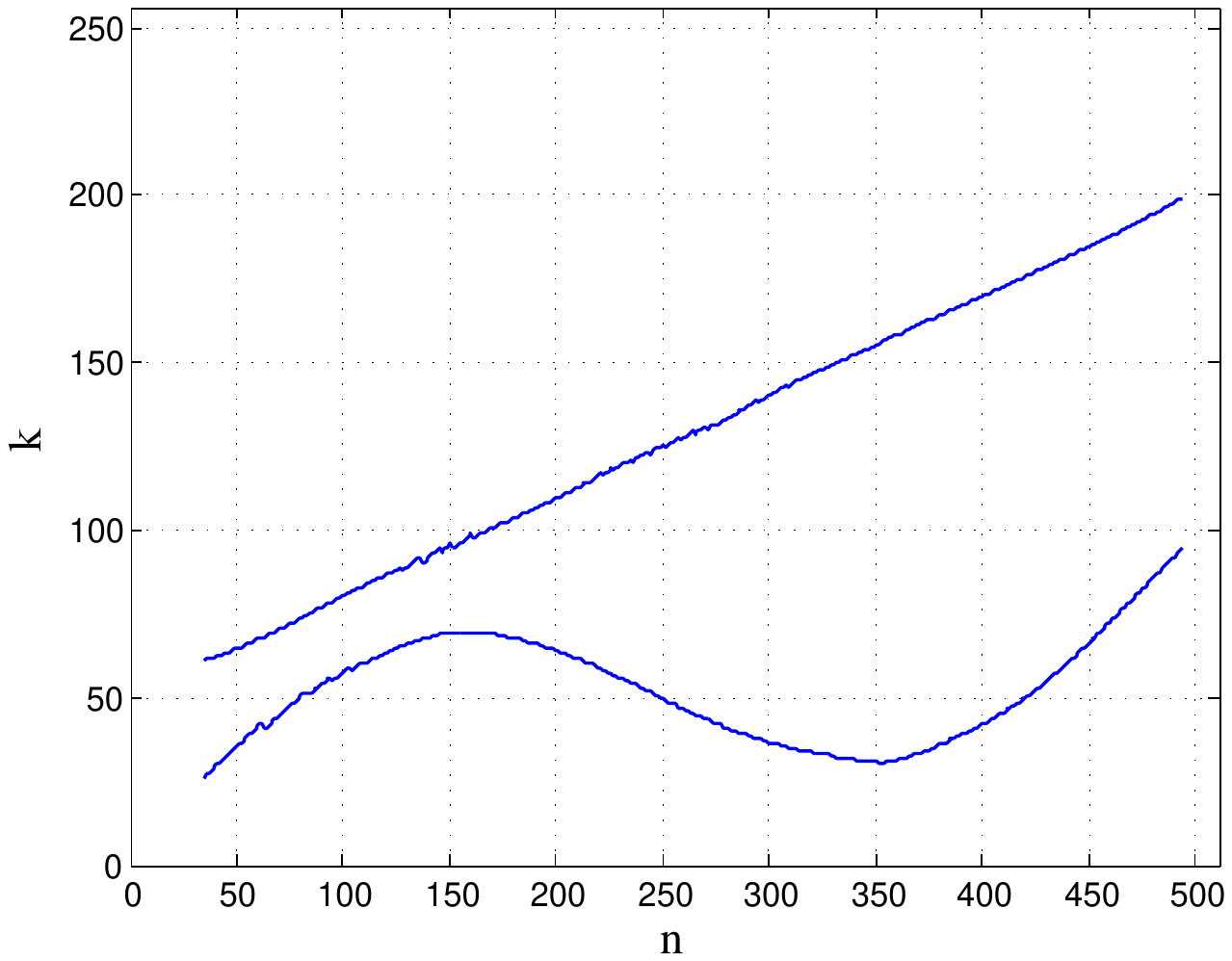}\\
  \vspace{-0.5cm}
  \centerline{(e)}
  \end{minipage}%
   \begin{minipage}[b]{0.5\linewidth}
  \vspace{-0.5cm}
  \centering
  \includegraphics[trim= 3cm 7cm 3cm 6.5cm, clip, width=5cm]{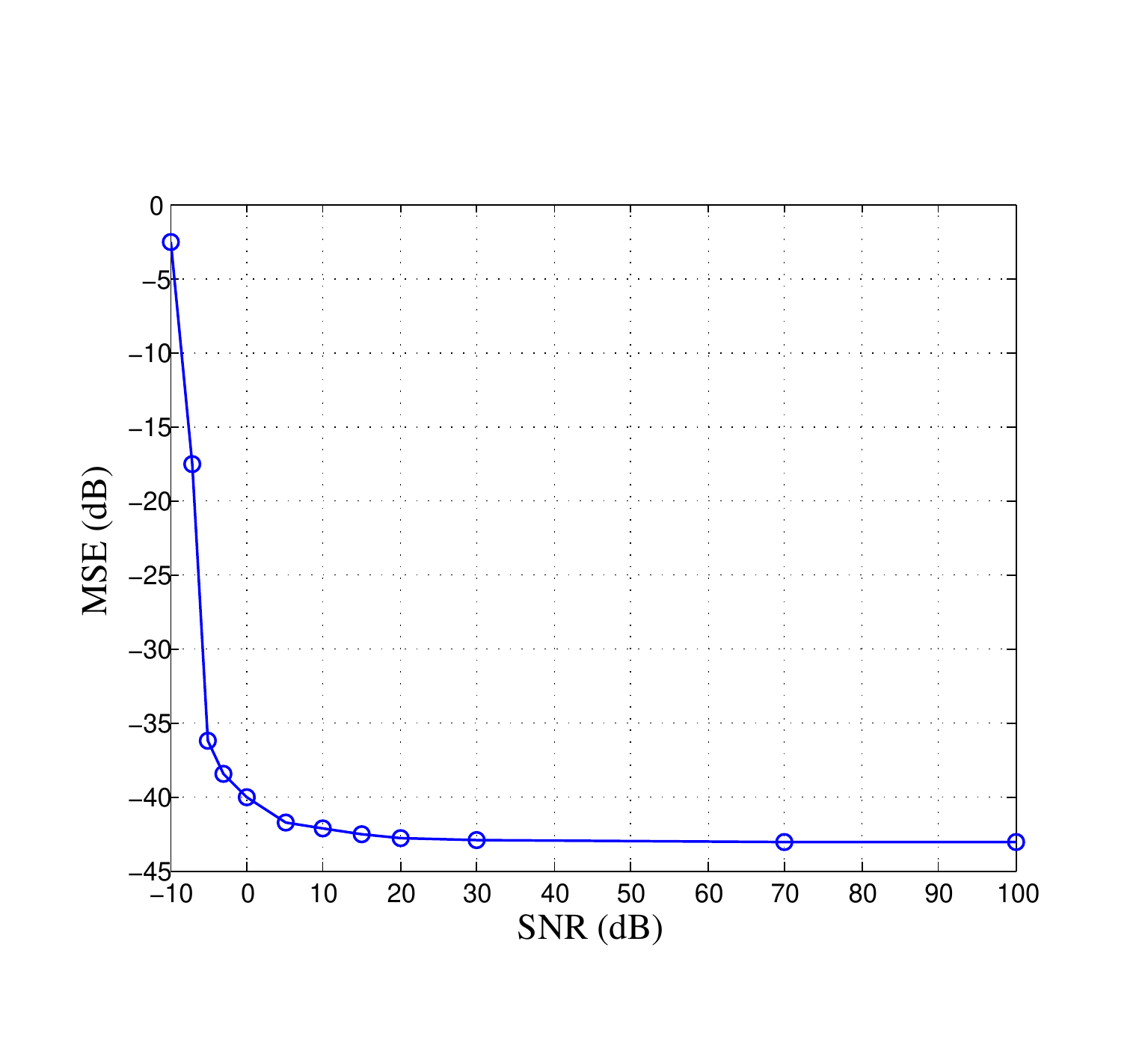}\\
  \vspace{-0.5cm}
  \centerline{(f)}
 \end{minipage}
\caption{Example 1 $x_1(n)$ with SNR $=-5$ dB:  (a) Wigner distribution of $x_1(n)$; (b) STFT with hamming window of length $64$; (c) Synthesized Wigner distribution; (d) original IF; (e) Estimated IF; (f) Mean-square error.}
\label{fig2}
\end{figure}

\medskip
\noindent
{\bf Example 1.} Consider the multicomponent signal $x_1(n)$
\begin{eqnarray*}
x_1(n)&=&\exp\left(j\frac{\pi}{256}(0.15n^2+50n)\right) \\
&&+\exp\left(j\left(\frac{\pi}{256}0.1n^2-40\cos(\frac{\pi}{500}n)\right)\right)+\mathcal{N}(n)
\end{eqnarray*}
 where $\mathcal{N}(n)$ is a complex white gaussian noise with a total of variance $\sigma^2$ is added to the signal. Figures \ref{fig2} (a) and (b) display the WD and the short time Fourier transform (STFT) of $x_1(n)$ for a SNR$=-5$ dB while Fig. \ref{fig2} (c) shows the superposition of the WDs of the chirp components (synthesized WD).  Notice that the WD does not clearly display the chirps due to cross-components and the smearing of the noise over the time-frequency space and the STFT is not robust against noise. The estimated and the original instantaneous frequencies of the signal $x_1(n)$ at SNR$=-5$ dB are given in Figs. \ref{fig2} (d) and (e). The mean square error (MSE) for the instantaneous frequency is shown in Fig. \ref{fig2} (f). It shows that the estimated IF using the proposed method matches well the original IF even at low SNRs.

\begin{table}[h]
\renewcommand{\arraystretch}{1.3}
\caption{Comparison of mean square error (MSE) for different time-frequency distributions with four different SNRs }
\label{table1}
\centering
\begin{tabular}{c|c|c|c|c}
\hline
\bfseries Time-frequency & \multicolumn{3}{c} \bfseries SNR (dB)\\
\bfseries Distribution  \\ \cline{2-5}
    & -5 & 0 & 5 & 100 \\
\hline \hline
Synthesized WD & -36.2 dB & -40.01 dB & -41.78 dB & -43.06 dB \\
STFT & -5.96 dB & -34.42 dB & -38.79 dB & -42.01 dB \\
WD & -3.93 dB & -5.47 dB & -6.72 dB & -6.95 dB \\
\hline
\end{tabular}
\end{table}

\begin{figure}[h]
\begin{minipage}[b]{0.5\linewidth}
\vspace{-0.5cm}
  \centering
 \includegraphics[trim= 3cm 7cm 3cm 6.5cm, clip, width=5cm]{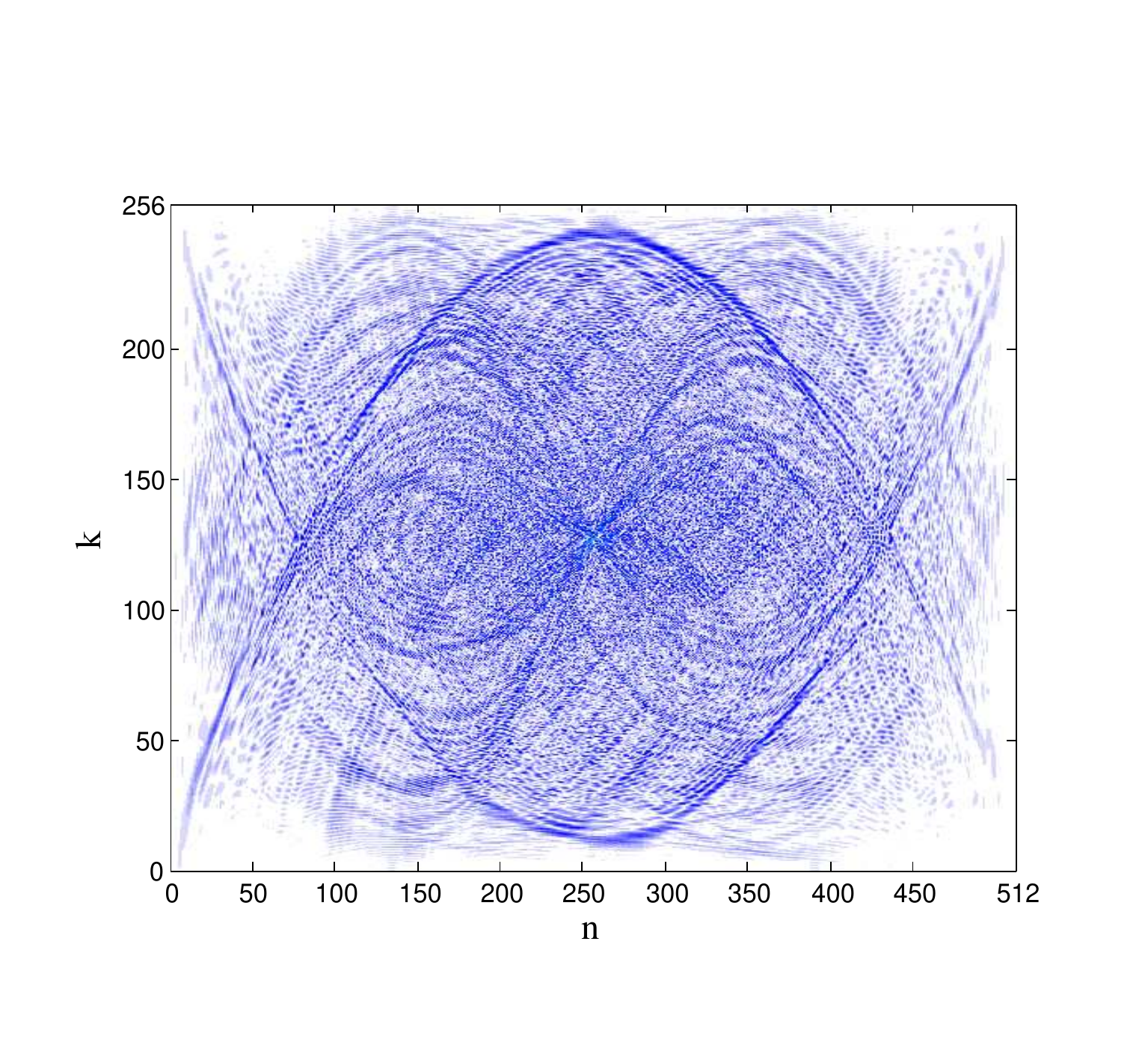}\\ 
 \vspace{-0.5cm}
  \centerline{(a)}
  \end{minipage}%
  \begin{minipage}[b]{0.5\linewidth}
  \vspace{-0.5cm}
  \centering
  \includegraphics[trim= 3cm 7cm 3cm 6.5cm, clip, width=5cm]{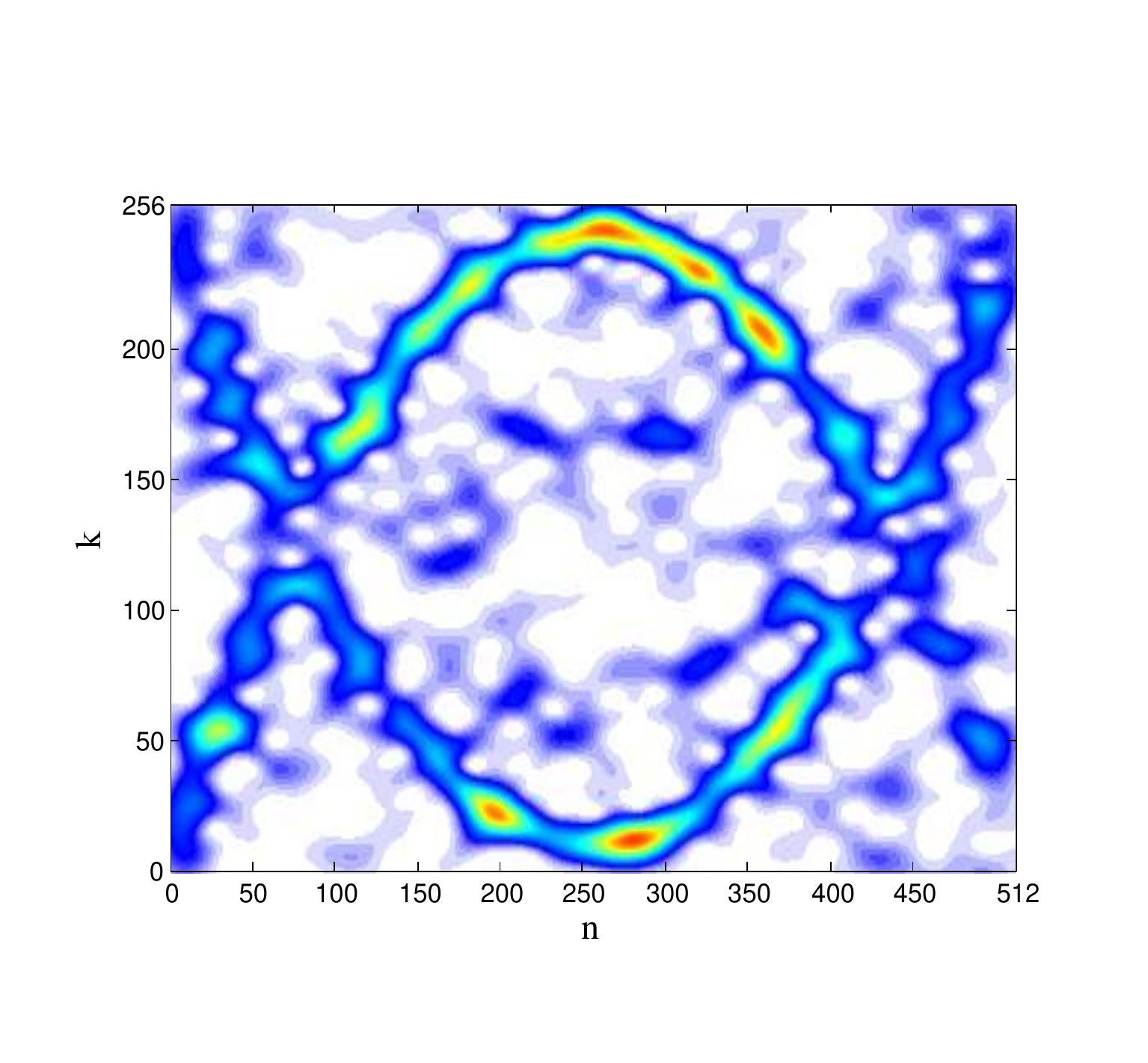}\\
  \vspace{-0.5cm}
  \centerline{(b)}
 \end{minipage}
 \begin{minipage}{0.5\linewidth}
 \vspace{-0.5cm}
\centering
\includegraphics[trim= 3cm 7cm 3cm 6.5cm, clip, width=5cm]{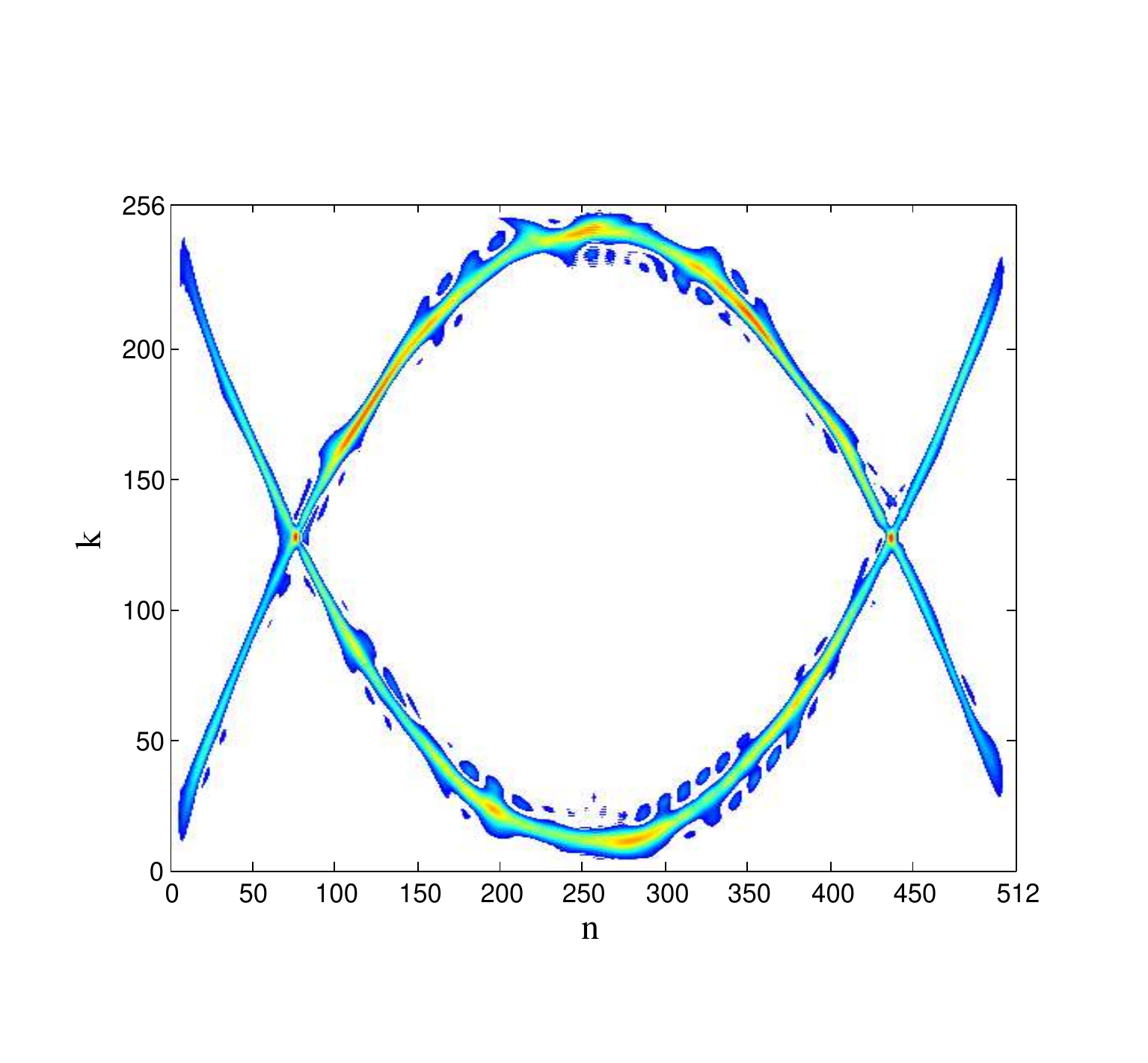}\\
\vspace{-0.5cm}
\centerline{(c)}
\end{minipage}%
\begin{minipage}{0.5\linewidth}
\vspace{-0.5cm}
\centering
\includegraphics[trim= 3cm 7cm 3cm 6.5cm, clip, width=5cm]{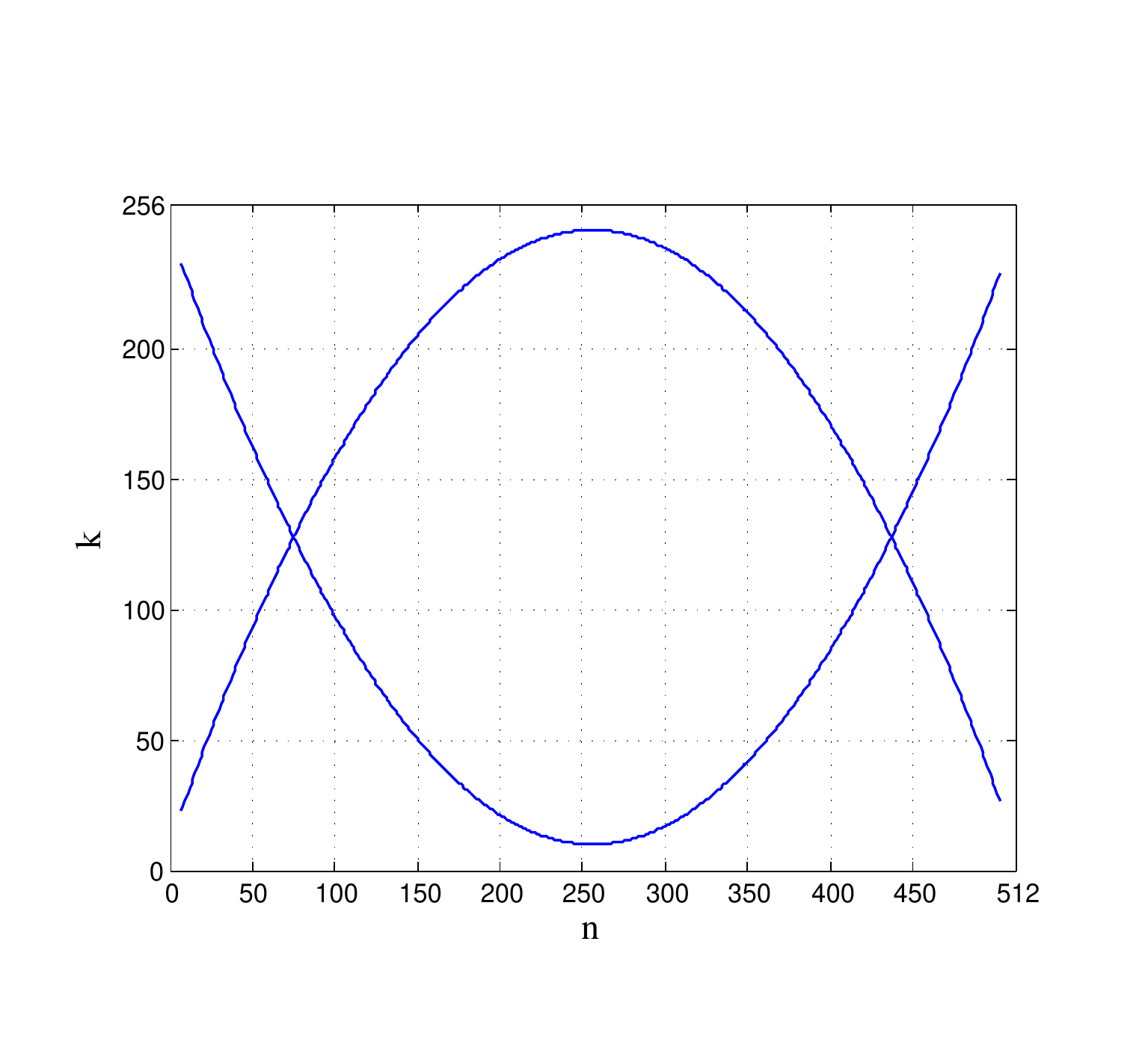}\\
\vspace{-0.5cm}
\centerline{(d)}
\end{minipage}
 \begin{minipage}[b]{0.5\linewidth}
  \vspace{-0.5cm}
  \centering
  \includegraphics[trim= 3cm 7cm 3cm 6.5cm, clip, width=5cm]{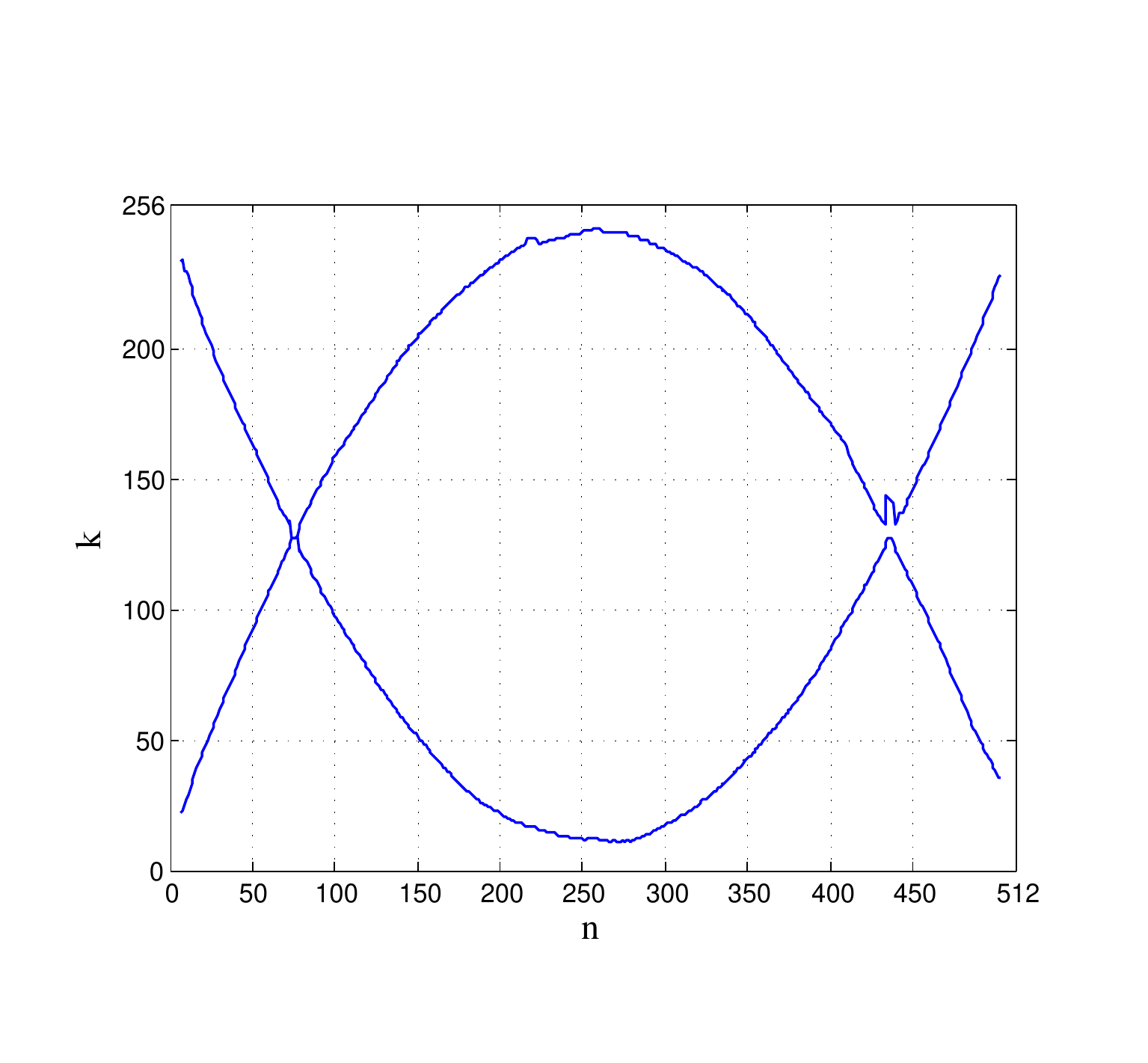}\\
  \vspace{-0.5cm}
  \centerline{(e)}
  \end{minipage}%
   \begin{minipage}[b]{0.5\linewidth}
  \vspace{-0.5cm}
  \centering
  \includegraphics[trim= 3cm 7cm 3cm 6.5cm, clip, width=5cm]{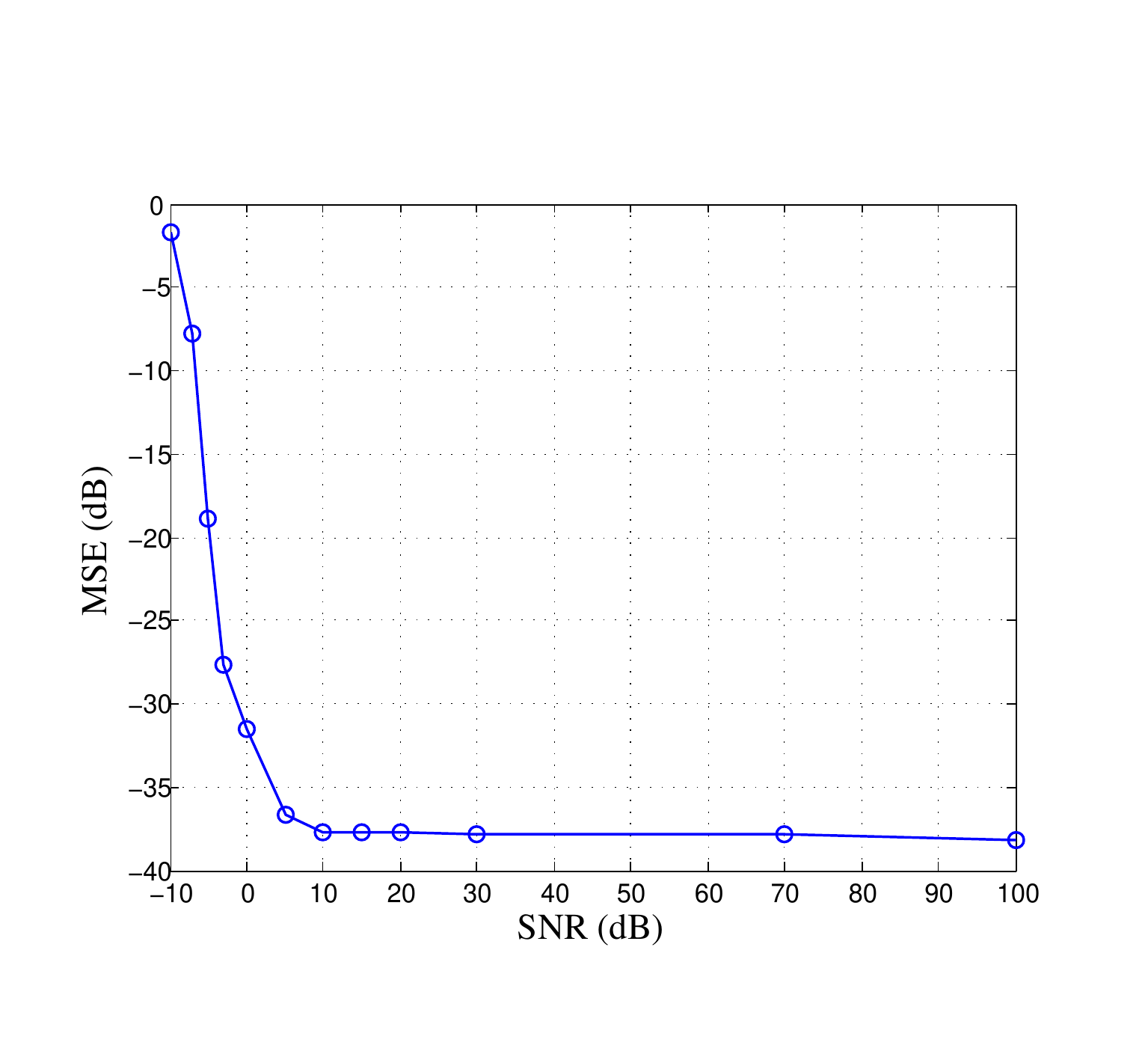}\\
  \vspace{-0.5cm}
  \centerline{(f)}
 \end{minipage}
\caption{Example 2 $x_2(n)$ with SNR $=0$ dB:  (a) Wigner distribution of $x_2(n)$; (b) STFT with hamming window of length $64$; (c) Synthesized Wigner distribution; (d) original IF; (e) Estimated IF; (f) Mean-square error.}
\label{fig3}
\end{figure}

\medskip
\noindent
{\bf Example 2.} Let the signal $x_2(n)$ be a multicomponent signal which has two intersected components in the time-frequency plane. The considered signal is embedded in noise as
\begin{eqnarray*}
x_2(n)&=&\exp\left(j\frac{\pi}{256}(\xi(n-256)^3+10n)\right) \\
&&+\exp\left(j\frac{\pi}{256}(\xi(n-256)^3-246n)\right)+\mathcal{N}(n)
\end{eqnarray*}
where $\xi=4\times10^{-4}$. The WD, STFT, and synthesized WD of the signal $x_2(n)$ with SNR$=0$ dB are shown in Figs. \ref{fig3} (a), (b), and (c). Figures \ref{fig3} (d) and (e) illustrate the original IF($\omega(n)$) as well as its  estimate ($\hat{\omega}(n)$). The MSE error as a function of SNR is given in Fig. \ref{fig3} (f).

\begin{table}[h]
\renewcommand{\arraystretch}{1.3}
\caption{Comparison of mean square error (MSE) for different time-frequency distributions with four different SNRs }
\label{table2}
\centering
\begin{tabular}{c|c|c|c|c}
\hline
\bfseries Time-frequency & \multicolumn{3}{c} \bfseries SNR (dB)\\
\bfseries Distribution  \\ \cline{2-5}
    & -5 & 0 & 5 & 100 \\
\hline \hline
Synthesized WD & -18.85 dB & -31.56 dB & -34.96 dB & -38.14 dB \\
STFT & -4.12 dB & -13.26 dB & -20.22 dB & -23.28 dB \\
WD & -0.74 dB & -0.91 dB & -1.36 dB & -1.88 dB \\
\hline
\end{tabular}
\end{table}

Tables \ref{table1} and \ref{table2} summarize the MSE measured in dB for the estimated IF using synthesized WD, STFT, and WD under the effect of noise. They show the synthesized WD is more robust against noise attack and gives better IF estimation than the other time-frequency distributions. On the other hand, the WD presents poor IF estimate even for high SNRs because it suffers from cross terms interference. The STFT shows good results for high SNRs but it gives poor IF estimate for low SNRs.

\begin{figure}[h]
\begin{minipage}[b]{0.5\linewidth}
\vspace{-0.5cm}
  \centering
 \includegraphics[trim= 3cm 7cm 3cm 6.5cm, clip, width=5cm]{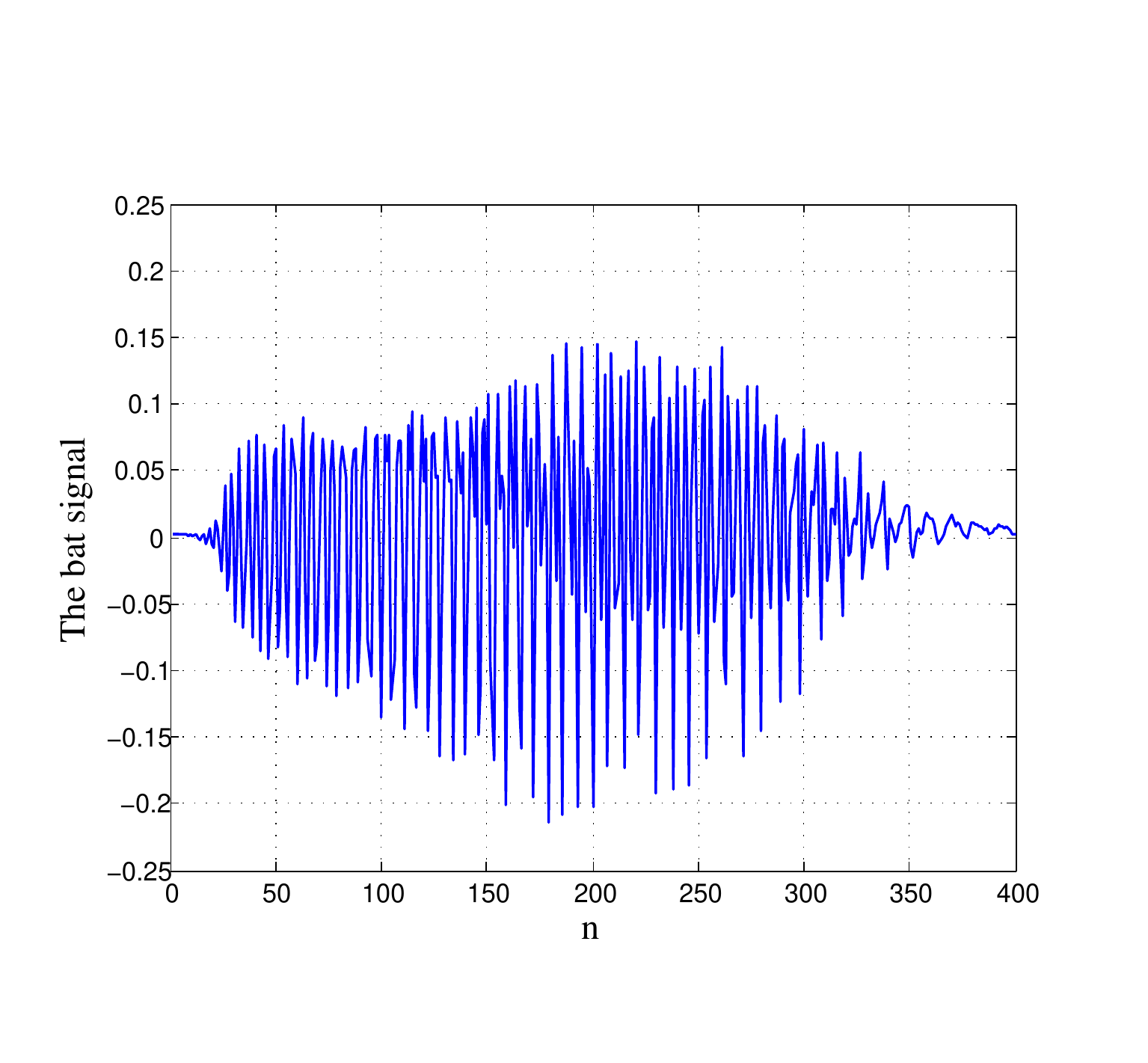}\\
 \vspace{-0.5cm}
  \centerline{(a)}
  \end{minipage}%
  \begin{minipage}[b]{0.5\linewidth}
  \vspace{-0.5cm}
  \centering
  \includegraphics[trim= 3cm 7cm 3cm 6.5cm, clip, width=5cm]{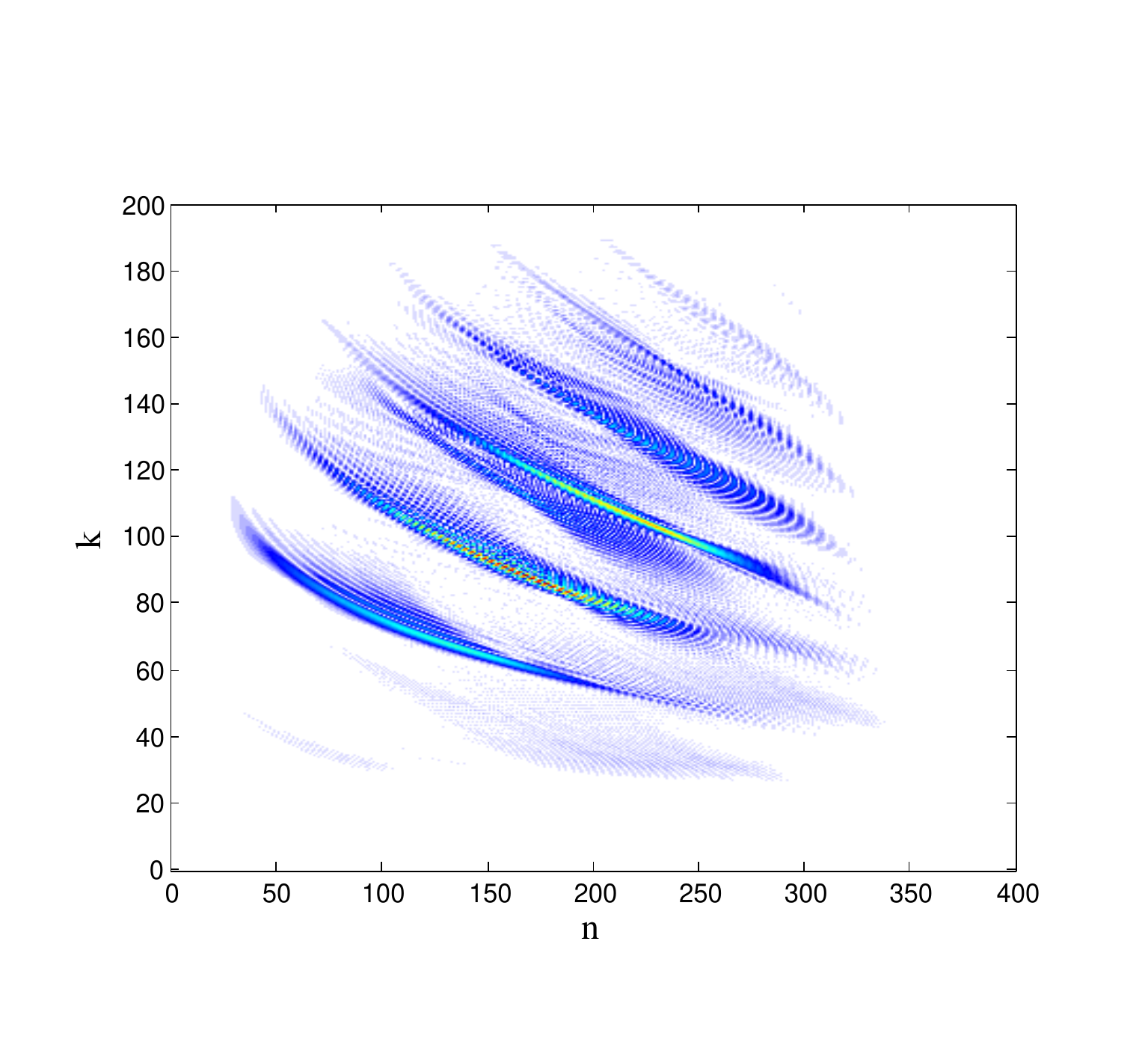}\\
  \vspace{-0.5cm}
  \centerline{(b)}
 \end{minipage}
 \begin{minipage}{0.5\linewidth}
 \vspace{-0.5cm}
\centering
\includegraphics[trim= 3cm 7cm 3cm 6.5cm, clip, width=5cm]{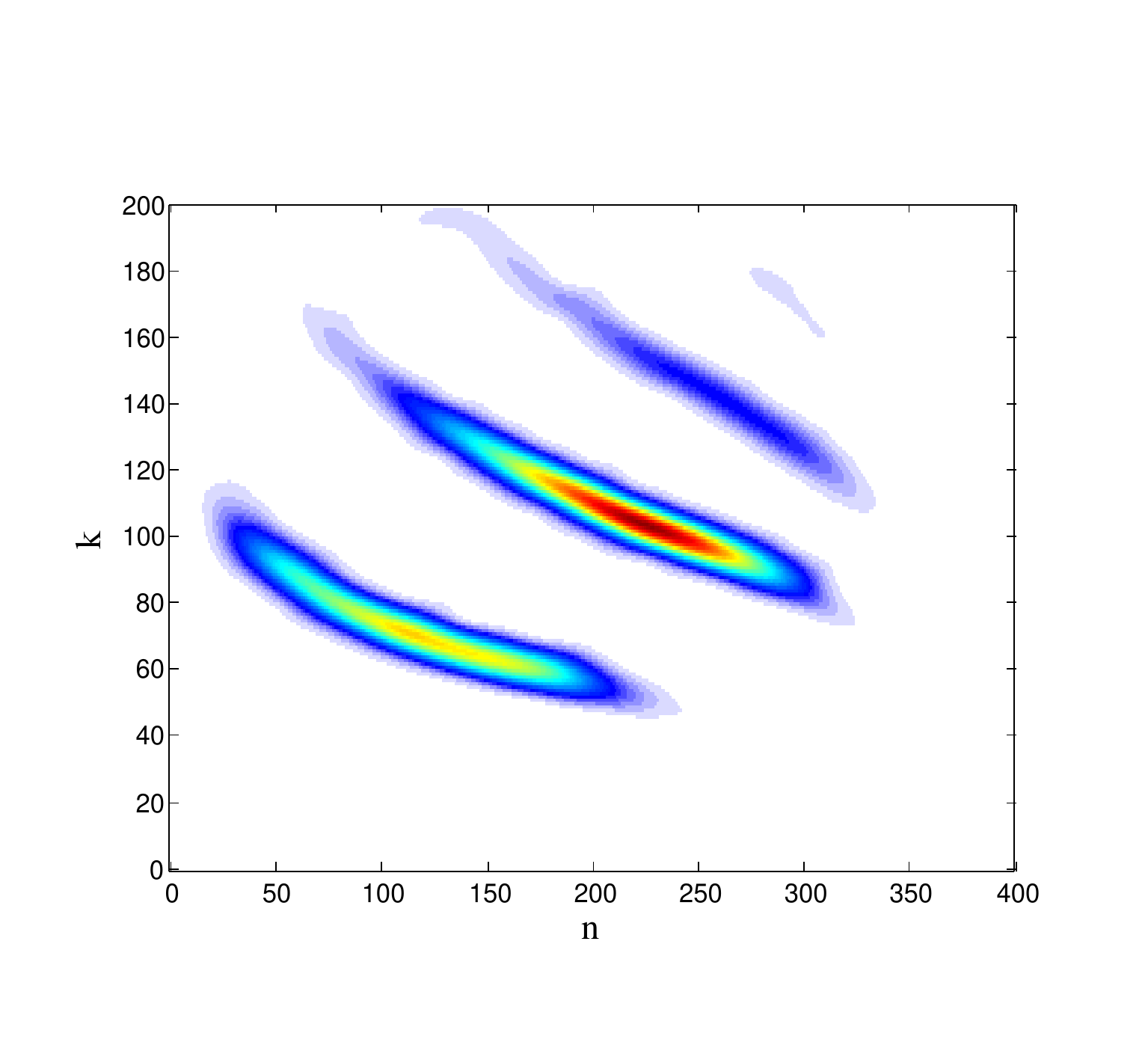}\\
\vspace{-0.5cm}
\centerline{(c)}
\end{minipage}%
\begin{minipage}{0.5\linewidth}
\vspace{-0.5cm}
\centering
\includegraphics[trim= 3cm 7cm 3cm 6.5cm, clip, width=5cm]{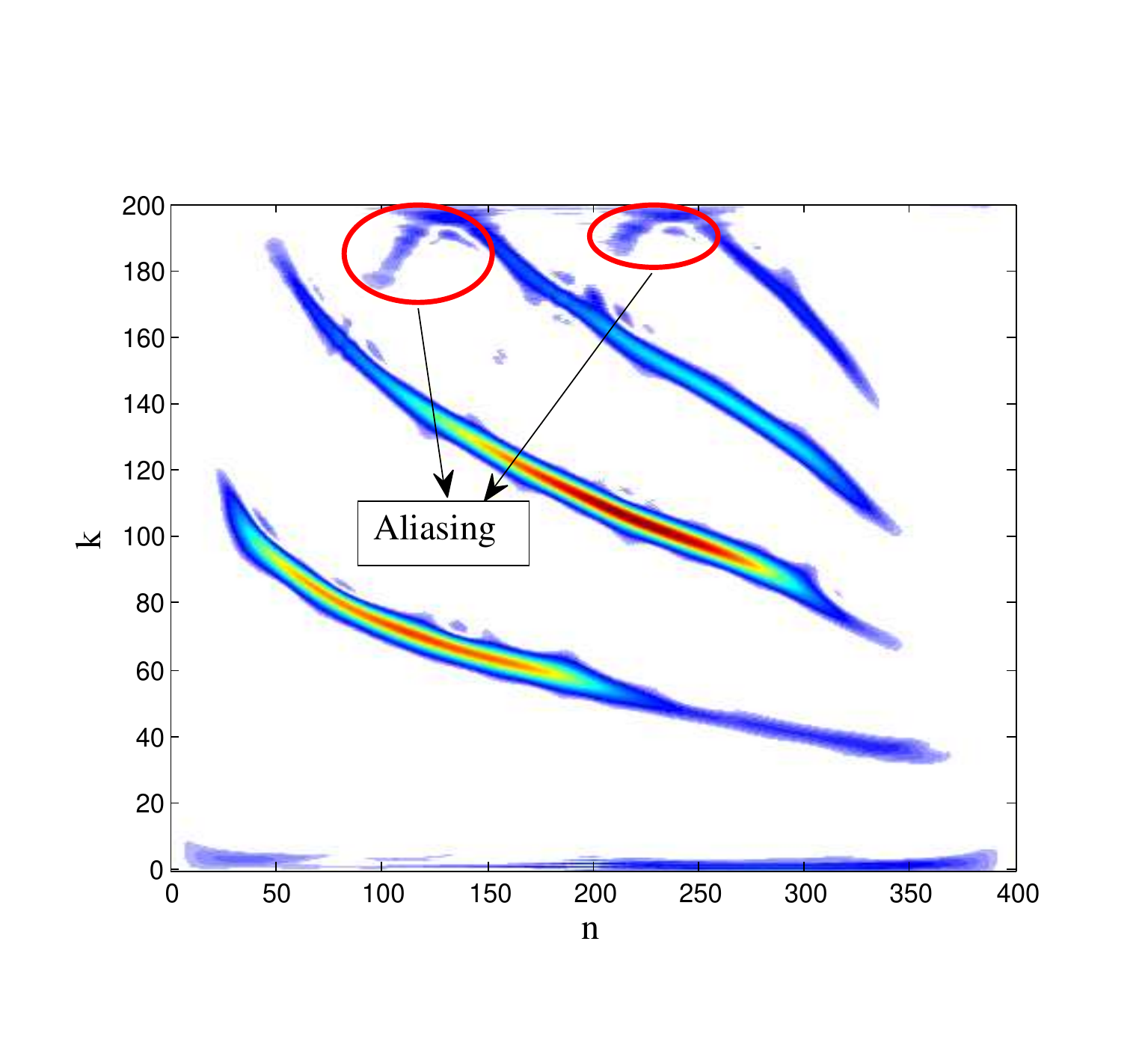}\\
\vspace{-0.5cm}
\centerline{(d)}
\end{minipage}
 \begin{minipage}[b]{0.5\linewidth}
  \vspace{-0.5cm}
  \centering
  \includegraphics[trim= 3cm 7cm 3cm 6.5cm, clip, width=5cm]{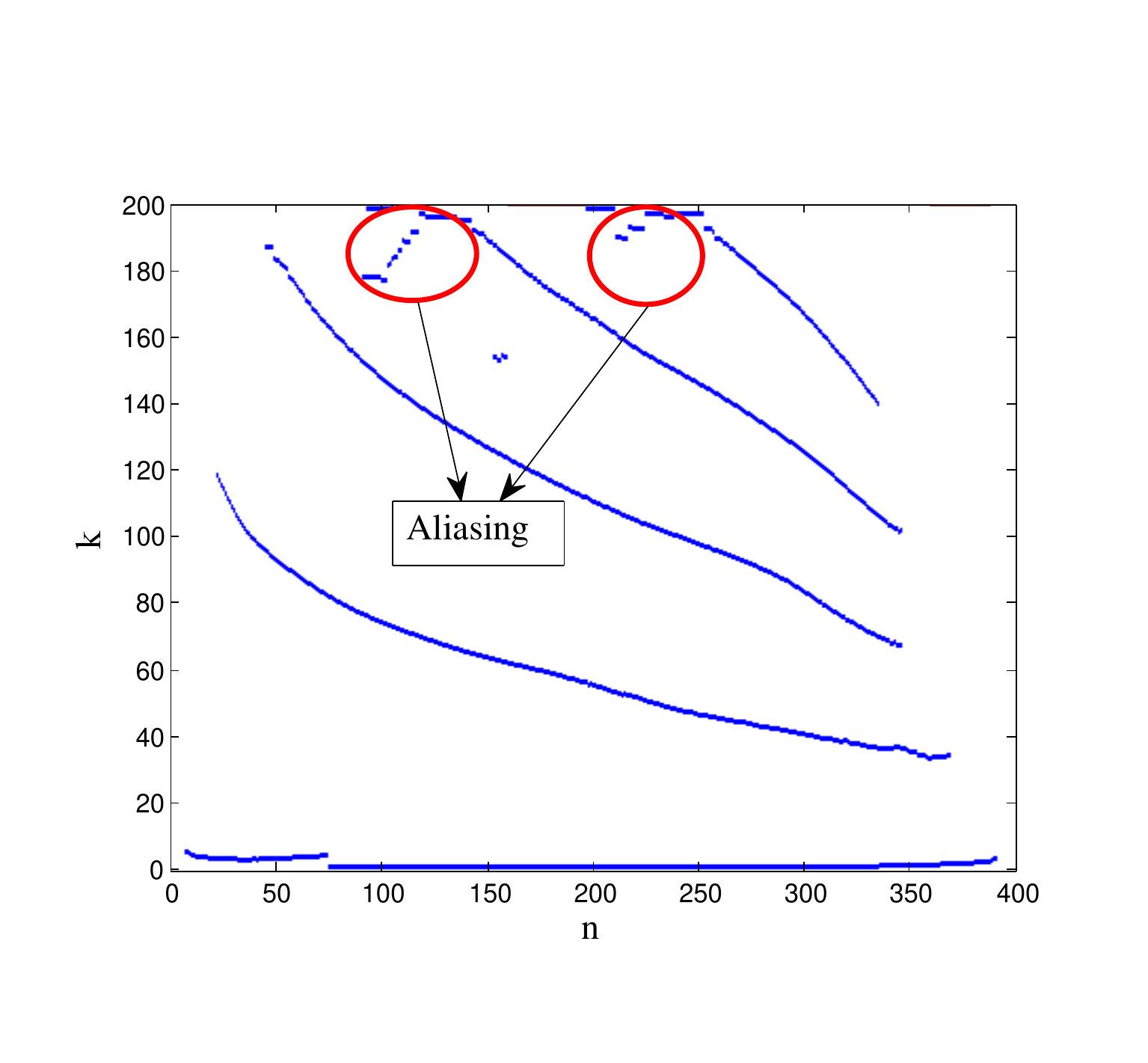}\\
  \vspace{-0.5cm}
  \centerline{(e)}
  \end{minipage}
\caption{Example 2:  (a) The bat signal in the time domain; (b) Wigner distribution of the bat signal; (c) STFT with hamming window of length $64$; (d) Synthesized Wigner distribution; (e) Estimated IF.}
\label{fig4}
\end{figure}

\medskip
\noindent {\bf Example 3.} In this example we show the potential of our algorithm in estimating the IF of actual multicomponent signals such as bat echolocation signal. The signal is given in Fig. \ref{fig4} (a). The WD and the STFT of it are given in Figs. \ref{fig4} (b) and (c) whereas Fig. \ref{fig4} (d) shows the synthesized WD using the proposed method. The estimated IF of the bat signal is illustrated in Fig. \ref{fig4} (e). The proposed IF estimation method performs well since it shows five components in the time frequency plane. In addition, we can observe that the bat signal  suffers from aliasing in the third and fourth components as explained in Figs. \ref{fig4} (d) and (e). Comparing our results with \cite{Lerga11} for the same bat signal, our IF algorithm gives better estimation because it shows more information of the signal in the time-frequency plane.

\section{Conclusions}
In this paper, we propose a new method for IF estimation based on the discrete linear chirp transform and the Wigner distribution. It is shown, that we can approximate locally a signal by linear chirps using the DLCT.  Separating them, finding the WD of each of these linear chirp and superposing them, a WD free of cross-terms is obtained for the signal under analysis. Simulations show we can obtain accurate IF estimation by the proposed method for even low levels of SNRs. Our procedure takes advantage of the maximum energy concentration of the Wigner distribution of linear chirps obtained from the DLCT.  Work is underway on the application of this procedure in biomedical applications.



\section*{Acknowledgment}
The authors wish to thank Curtis Condon, Ken White, and Al Feng of the Beckman Institute of the University of Illinois for the bat data and for permission to use it in this paper.












\end{document}